\documentclass[letterpaper,11pt]{article}

\usepackage{mathrsfs}
\usepackage{epsf}
\usepackage{amsmath,amsfonts}
\usepackage{graphicx}
\usepackage{latexsym}

\def\prob{\mbox{Pr}}

\newtheorem{theorem}{Theorem}

\newtheorem{definition}[theorem]{Definition}

\begin{document}

\baselineskip=20pt

\begin{center}

{\Large\bf On Simulating Nondeterministic 
Stochastic Activity Networks}\footnote{This work is partially supported by CNPq, CAPES, and a FAPERJ BBP grant.}

Valmir C. Barbosa\footnote{COPPE/PESC,
UFRJ, C.Postal 68511, Rio de Janeiro, Brazil, CEP 21941-972.
E-mail: {\tt valmir@cos.ufrj.br}}
\hspace{0.8cm}        
Fernando M. L. Ferreira\footnote{NCE, UFRJ, 
C.Postal 2324, Rio de Janeiro, Brazil, CEP 20001-970. 
E-mail: {\tt fmachado@nce.ufrj.br}}
\hspace{0.8cm}
Daniel V. Kling\footnote{NCE, UFRJ,
C.Postal 2324, Rio de Janeiro, Brazil, CEP 20001-970.
E-mail: {\tt danielvk@posgrad.nce.ufrj.br}}
\hspace{0.8cm}

Eduardo Lopes\footnote{NCE, UFRJ, 
C.Postal 2324, Rio de Janeiro, Brazil, CEP 20001-970. 
E-mail: {\tt eduardolopes@gmx.net}}
\hspace{0.8cm}
F\'abio Protti\footnote{IM and NCE, UFRJ, 
C.Postal 2324, Rio de Janeiro, Brazil, CEP 20001-970. 
E-mail: {\tt fabiop@nce.ufrj.br}}
\hspace{0.8cm}
Eber A. Schmitz\footnote{IM and NCE, UFRJ, 
C.Postal 2324, Rio de Janeiro, Brazil, CEP 20001-970. 
E-mail: {\tt eber@nce.ufrj.br}}

\end{center}

{\bf Abstract.} In this work we deal with a mechanism for process simulation called a {\em NonDeterministic Stochastic Activity Network} (NDSAN). An NDSAN consists basically of a set of {\em activities} along with {\em precedence relations} involving these activities, which determine their order of execution. Activity durations are stochastic, given by continuous, nonnegative random variables. The nondeterministic behavior of an NDSAN is based on two additional possibilities: (i) by associating choice probabilities with groups of activities, some branches of execution may not be taken; (ii) by allowing iterated executions of groups of activities according to predetermined probabilities, the number of times an activity must be executed is not determined {\em a priori}. 
These properties lead to a rich variety of activity networks, capable of modeling many real situations in process engineering, project design, and troubleshooting.
We describe a recursive simulation algorithm for NDSANs, whose repeated execution produces a  close approximation to the probability distribution of the completion time of the entire network. We also report on real-world case studies.       

{\bf Keywords:} activity networks, stochastic activity networks, nondeterministic activity networks, stochastic project scheduling problems. 

\section{Introduction}

In this work we deal with a mechanism for process simulation called a {\em NonDeterministic Stochastic Activity Network} (NDSAN). An NDSAN consists basically of a set of {\em activities} along with {\em precedence relations} involving these activities, which determine their order of execution. This order is captured by a digraph with some special properties: the possibility of defining {\em nondeterministic branches of execution}, by associating choice probabilities with some activities, and {\em loops of execution}, which specify the iterated execution of a group of activities according to predetermined loop probabilities. These properties allow for a rich variety of activity networks, capable of modeling many real situations in process engineering, project design, and troubleshooting. 

There are two main types of activity networks. A {\em deterministic activity network} is represented by a precedence digraph whose topology remains {\em fixed} as the activities are executed. Examples of deterministic activity networks include CPM and PERT networks, see e.g.~\cite{MPD83}. On the other hand, a nondeterministic activity network allows for the possibility of a dynamic topology. Examples of such networks are inhomogeneous Markov chains, GANs (Generalized Activity Networks)~\cite{E77}, and GERT (Graphical Evaluation and Review Technique) networks~\cite{MC76}.

The duration of each network activity is given by a random variable. Thus, a fundamental problem is determining the distribution of the completion time of the entire network. For deterministic activity networks, this general problem is known as the Stochastic Project Scheduling Problem~\cite{DH02}. 

Our definition of NDSANs \ combines stochastic activity durations with nondeterminism. In an NDSAN, activities are represented by {\em nodes}, and an arc oriented from activity $a_i$ to activity $a_j$ means that the execution of $a_j$ may only start after the execution of $a_i$ has ended. Nondeterminism is achieved, as indicated above, by means of two possibilities: (i) some branches of execution are not necessarily taken, and (ii) the number of times a group of activities is to be executed is not determined {\em a priori}. These additional possibilities are supported by the introduction of two new categories of nodes, namely {\em decision nodes} and {\em loop nodes}. A decision node associates {\em probabilities} with its out-neighbors and selects one of them to be executed accordingly; this selection is interpreted as one possible deterministic scenario among many. A loop node allows the repeated execution of a group of activities, the number of iterations depending on probabilities associated with the loop node. Loop nodes are particularly interesting to model refinement processes, such as quality control and error testing/correction. We also define {\em junction nodes} for adequately combining the two new constructions into the network. In Section~2 we define NDSANs formally, in terms of recursive construction steps that combine smaller NDSANs into larger ones via certain types of structured templates. 

In Section~3, we give an analytical description of the random variable $T[D]$ associated with the completion time of NDSAN $D$. We assume that the duration of each activity $a_i$ in $D$ is given by a continuous, nonnegative random variable $T_i$. The random variable $T[D]$ is thus given in terms of the $T_i$'s and the probabilities associated with the decision/loop nodes.

Although $T[D]$ can be described precisely, we lack a closed-form expression for it and even numerical methods to find its distribution from such a description may be computationally too hard, especially when the number of activities is large. In Section~4, we describe a recursive simulation algorithm whose execution returns a single plausible value (``observation'') in the sample space of $T[D]$. Running the simulation algorithm a suitable number $N$ of times produces a close approximation to the probability distribution of $T[D]$. The value of $N$ can be obtained by using the same statistic as the Kolmogorov-Smirnov test, see e.g.~\cite{Hoel} (Section~13.5), as we also discuss in Section~4. 
 
Section~5 presents two computational experiments. For each experiment, the result of the simulations is shown as a frequency histogram together with a fitting curve that approximates the expected shape of the density of $T[D]$, an approximate probability distribution of $T[D]$, and an approximate probability density of $T[D]$ obtained from the approximate distribution. Section~6 discusses ongoing work. 

In a recent related work, Leemis {\em et al.}~\cite{LDDMC06} develop algorithms to calculate the probability distribution of the completion time of a stochastic activity network with continuous activity durations. In their work, activities are modeled by {\em arcs} and the networks are acyclic and deterministic (i.e., allow no variation in topology). The authors describe a recursive Monte Carlo simulation algorithm, which is network-specific and must therefore be rewritten specifically for each new network. Also, they provide two exact algorithms, one for series-parallel networks and another for more general networks whose nodes have at most two incoming arcs each.

We remark that all the discussion on random variables in this work can be adapted to the 
case of discrete random variables. (In~\cite{shier}, pp.\ 122--123, for example, an activity network with discrete activity durations is given.)

\section{Formal definition of NDSANs}

In this work, $D$ denotes a digraph with $n$ nodes and $m$ arcs.
If $(v,w)$ is an arc of $D$, then node $v$ 
is an {\em in-neighbor} of node $w$, whereas $w$ 
is an {\em out-neighbor} of $v$. By disregarding arc orientation, 
we may also simply say that $v$ and $w$ are {\em neighbors}. A 
node having no in-neighbors (resp. out-neighbors) 
is called a {\em source node} (resp. {\em sink node}). 
If $D$ is a digraph containing a single source (resp. sink) node $v$ , 
then $v$ is denoted by $source(D)$ (resp. $sink(D)$). 

An NDSAN is a special digraph whose node set is partitioned into 
four subsets of nodes: a subset $S_a=\{a_i \mid 1 \leq i \leq n_a\}$ of 
{\em activity nodes}; a subset $S_b=$ $\{b_i \mid 1 \leq i \leq n_b\}$ of  
{\em junction nodes}; a subset $S_d=\{d_i \mid 1 \leq i \leq n_d\}$ of  
{\em decision nodes}; and a subset $S_{\ell}=\{\ell_i \mid 1 \leq i \leq n_{\ell}\}$ 
of {\em loop nodes}.

An {\em activity node} $a_i$ represents a single 
{\em activity} (or {\em task}) to be executed in 
the network. The execution of $a_i$ starts only  
after the execution of {\em all} of its in-neighbors has ended. 
When the execution of $a_i$ ends,  
{\em all} of its out-neighbors start executing simultaneously.  
Each activity node $a_i$ has a {\em duration}  
({\em execution time}) $T_i$, which is a continuous, nonnegative random variable. 
We assume that the execution time of an activity node 
does not depend on the execution time of any other activity node.
That is, the $T_i$'s are independent random variables.    
An activity node is represented by a circle. 
See Figure~1(a). 

A {\em junction node} $b_i$ is used for a 
syntactic purpose. It may have several in-neighbors, 
but it has a single out-neighbor $v$. 
When the execution of any in-neighbor of $b_i$ ends, 
the execution of $v$ is started immediately. 
In other words, $b_i$ acts simply as a ``connecting 
point'' of incoming arcs. 
A junction node is represented by a square. 
See Figure~1(b).

A {\em decision node} $d_i$ is used to select 
one particular branch of the execution flow, as described in what follows.
By construction, all of $d_i$'s neighbors are activity nodes.    
It has a single in-neighbor $a_h$  
and $\alpha_i \geq 2$ out-neighbors $a_{j_1}, \ldots, a_{j_{\alpha_i}}$. 
The execution of $d_i$ is assumed to be instantaneous,  
and consists of selecting exactly one of its out-neighbors, say $a_{j_k}$,  
as the next node to execute. The activity node $a_{j_k}$ 
is selected by $d_i$ with probability $p^i_k$, 
$k=1,\ldots,\alpha_i$, such that $\sum_{k=1}^{\alpha_i} p^i_k = 1$.
A decision node is represented by a lozenge. 
See Figure~1(c).

A {\em loop node} $\ell_i$ represents the usual 
iteration mechanism. 
By construction, $\ell_i$ has a single in-neighbor   
(a junction node $b_h$) and 
two out-neighbors (activity nodes $a_r$ and $a_j$). 
After the execution of $b_h$,  
a Boolean condition $E_i$ associated with $\ell_i$ 
is instantaneously tested:  
if $E_i$ is {\em false} 
then $a_r$ is executed next, 
otherwise $a_j$ is.
An array of real values 
associated with ${\ell}_i$   
gives the sequence $q^i_1, \ldots, q^i_{\beta_i}$ 
of probabilities corresponding to $\beta_i$ consecutive 
passages through $\ell_i$,  
in such a way that 
the probability that $E_i$ is {\em false} 
at the $k$th passage 
through ${\ell}_i$ is $q^i_k$.  
That is, the probability of exiting the loop at  
this point is $1-q^i_{k}$. 
We assume that $q^i_{\beta_i}=0$ in order 
to guarantee the termination of the loop 
in at most $\beta_i$ consecutive passages 
through ${\ell}_i$.
A loop node is represented by a filled lozenge.  
See Figure~1(d).

\begin{figure}[t]
\label{nodes}
\epsfxsize=8.5cm
\centerline{\epsfbox{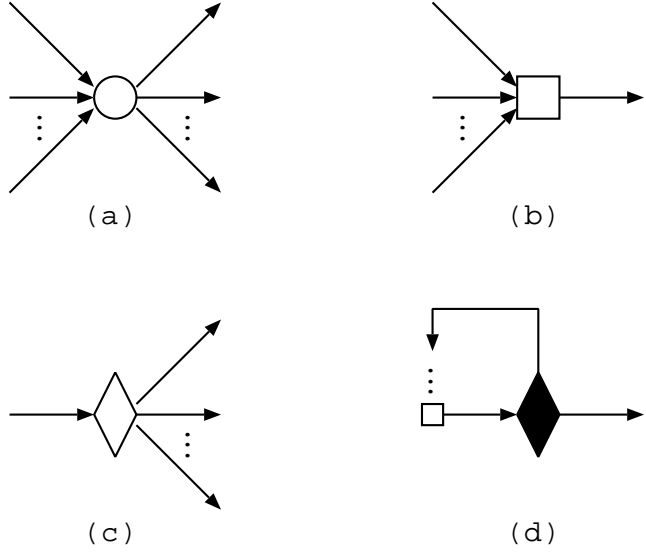}}
\caption{Types of node: (a) activity node; (b) junction node;
(c) decision node; (d) loop node.}
\end{figure}

We are now ready to give the formal definition of NDSANs 
in terms of recursive construction steps. 
The base NDSAN is a digraph consisting of a single activity node. 
In a general step, NDSANs containing a single source node 
and a single sink node are combined to yield a larger NDSAN. 

The recursive construction steps are based 
on the following {\it Substitution Rule:} 

\begin{description} 

\item {\bf Substitution Rule:} \ Let $D_0$ be a digraph and   
$\{v_1, v_2, \ldots, v_{\eta}\}$ a subset of its node set.  
Let $D_1, D_2, \ldots D_{\eta}$ be NDSANs, 
each containing a single source node and a single sink node. 
Construct an NDSAN $D$ by replacing $v_i$ by $D_i$, $1 \leq i \leq \eta$,  
in such a way that every input (output) arc of 
$v_i$ in $D_0$ is an input (output) arc of $source(D_i)$ ($sink(D_i)$) in $D$. 
Let $Sub(D_0, D_1, \ldots, D_{\eta})=D$.

\end{description}

\begin{definition} \label{NDSAN}

An NDSAN is defined as follows: 

\begin{enumerate}

\vspace{-0.5cm}

\item[$1.$] A digraph $D$ consisting of a single activity node 
is an NDSAN, called the {\em\bf trivial NDSAN}.

\item[$2.$] Let $D_1, D_2, \ldots D_{\eta}$ be NDSANs.  

\begin{enumerate}

\item[$2.1$] If $D_0$ is an acyclic digraph of node set 
$\{v_1, \ldots, v_{\eta}\}$
containing a single source node 
and a single sink node (Figure~2(a)), 
then $Sub(D_0, D_1, \ldots, D_{\eta})$ is an NDSAN, 
called an {\em\bf acyclic NDSAN} (Figure~2(b)). 

\item[$2.2$] If $D_0$ is the digraph in Figure~{\em 3(a)}, then  
$Sub(D_0, D_1, \ldots, D_{\eta})$ is an NDSAN, 
called a {\em\bf decision NDSAN} (Figure~3(b)).

\item[$2.3$] If $D_0$ is the digraph in Figure~{\em 4(a)}, 
then $Sub(D_0, D_1, D_2, D_3)$ is an NDSAN, 
called a {\em\bf loop NDSAN} (Figure~4(b)). 

\end{enumerate}

\end{enumerate}     

\end{definition}

\begin{figure}[t]
\label{acyc}
\epsfxsize=14cm
\centerline{\epsfbox{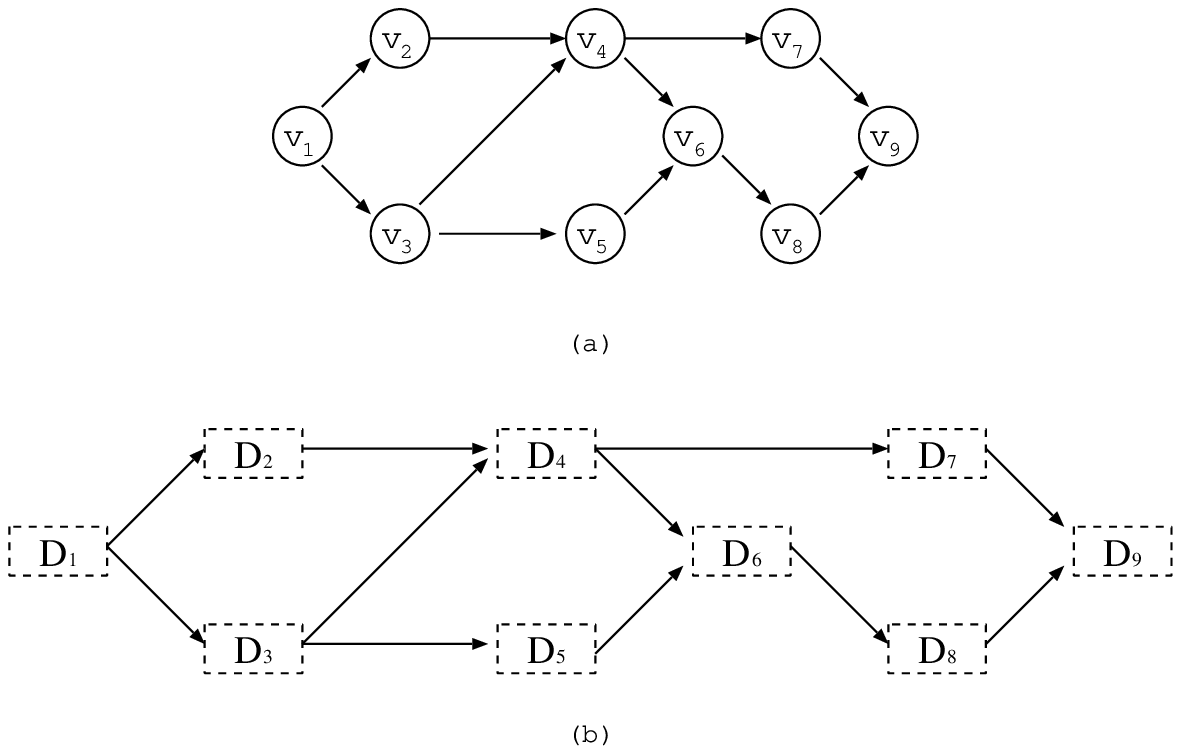}}
\caption{Construction of an acyclic NDSAN.}
\end{figure}

\begin{figure}[p]
\label{dec}
\epsfxsize=9cm
\centerline{\epsfbox{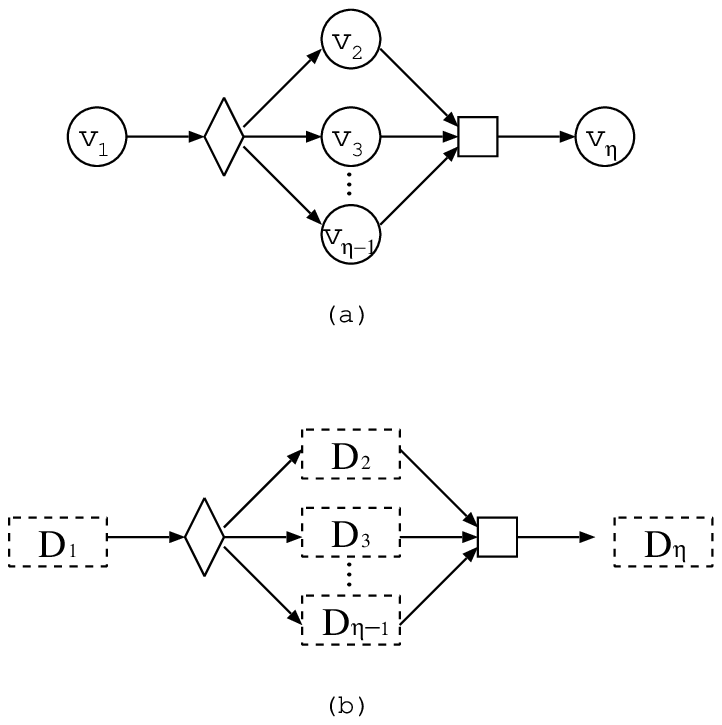}}
\caption{Construction of a decision NDSAN.}
\end{figure}

\begin{figure}[p]
\label{loop}
\epsfxsize=9cm
\centerline{\epsfbox{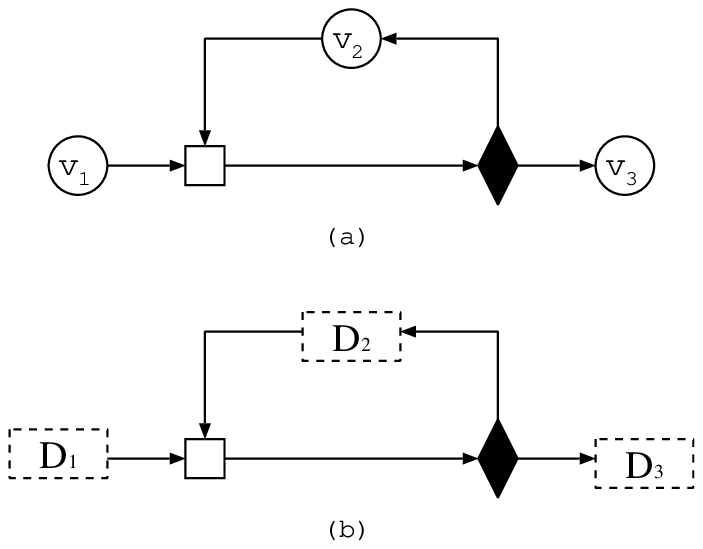}}
\caption{Construction of a loop NDSAN.}
\end{figure}

It is easy to see that the network $Sub(D_0, D_1, \ldots, D_{\eta})$ 
resulting from 2.1, 2.2, or 2.3 in the above definition contains a single  
source node and a single sink node, both activity nodes. 

{\bf Scope of the definition of NDSANs.} \ 
Although other definitions of NDSANs may be possible, 
we believe that Definition~\ref{NDSAN} not only determines a wide class  
of activity networks, but also allows the realization of any structured project, since it  
provides basic constructions that are generally thought to suffice for the specification of how 
concurrent tasks are to interrelate. In other words:

\begin{itemize}

\item[--] an acyclic NDSAN embodies the notion of multiple concurrent execution threads, 
which may be started as a single thread branches out into several independent ones, 
and terminated as they coalesce into a single thread for further execution. 

\item[--] a decision NDSAN allows for nondeterministic switches, or decision points, 
to be incorporated into the course of a thread's execution.

\item[--] a loop NDSAN allows any of the above to be iterated, possibly for a 
probabilistically selected number of times. 

\end{itemize}

\section{Execution time of an NDSAN}

In this section we use the following terminology and notation. 
(See, for instance,~\cite{Feller,Hoel}.)  
If $X$ is a random variable, then $F_X$ denotes the 
{\em probability distribution function} ({\em PDF}\,) of $X$, 
and $f_X$ the {\em probability density function} ({\it pdf}\,)   
of $X$. Recall that, for any $t$ in the domain of $X$, $F_X(t) = \prob(X \leq t)$.
If $X$ is a continuous variable, we have 
\begin{equation} \label{rel}
F_X(t) = \int_{-\infty}^t f_X(x) \ dx.
\end{equation}
Hereafter, the random variable standing for 
the execution time of NDSAN $D$
will be denoted by $T[D]$. This random variable 
can be determined as follows.




\bigskip

{\em Case 1: $D$ is a trivial NDSAN}

\bigskip 

Assuming that $D$ consists of the activity node $a_i$, 
we have $T[D]=T_i$.

\bigskip

{\em Case 2: $D$ is not a trivial NDSAN}

\bigskip

By 2.1, 2.2, and 2.3 in Definition \ref{NDSAN}, 
$T[D]$ can be recursively determined in terms of 
$T[D_1],T[D_2], \ldots,$ $T[D_{\eta}]$.  


\bigskip

{\em Case 2.1: $D$ is an acyclic NDSAN}  

\bigskip

Consider item 2.1 in Definition \ref{NDSAN}. 
Let ${\cal P}$ be the collection of all directed paths from 
$source(D_0)$ to $sink(D_0)$. Let $P \in {\cal P}$, and write  
$P=v_{i_1} v_{i_2} \ldots v_{i_{|P|}}$, 
where $|P|$ denotes the number of nodes of $P$. 
Let $D_{i_1}, D_{i_2}, \ldots, D_{i_{|P|}}$ be the  
NDSANs that substitute for $v_{i_1}, v_{i_2}, \ldots, v_{i_{|P|}}$. 
If $S_P$ is the time required for the serial execution of 
$D_{i_1}, D_{i_2}, \ldots, D_{i_{|P|}}$, then 
\begin{equation}
S_P = \sum_{k=1}^{|P|} T[D_{i_k}].
\end{equation}
(Recall that $T[D_{i_k}]$ is the random variable standing for the execution 
time of $D_{i_k}$, $1 \leq k \leq |P|$.) 

Since the $T[D_{i_k}]$'s are independent random 
variables, the pdf $f_{S_P}$ of $S_P$ is given by 
the convolution of the pdfs 
$f_{T[D_{i_1}]}, f_{T[D_{i_2}]}, \ldots, f_{T[D_{i_{|P|}}]}$, that is, 
\begin{equation} \label{fSP}
f_{S_P}(t) = (f_{T[D_{i_1}]} * f_{T[D_{i_2}]}* \cdots * f_{T[D_{i_{|P|}}]})(t).
\end{equation}
Define $f_1=f_{T[D_{i_1}]}$ and $f_k=f_{k-1} * f_{T[D_{i_k}]}$, $2 \leq k \leq |P|$. 
Then we have, for any $t$, 
\begin{equation}
f_k(t) = \int_0^{\infty} f_{k-1}(t-x)f_{T[D_{i_k}]}(x) \ dx
\ \ \mbox{ and } \ \ f_{S_P}(t) = f_{|P|}(t).
\end{equation}
Following Equation~(\ref{rel}), the PDF of $S_P$ is then given by 
\begin{equation} \label{FSP}
F_{S_P}(t) = \int_0^t f_{S_P}(x) \ dx.
\end{equation}

Having described the variables $S_P$ for $P \in {\cal P}$, 
the random variable $T[D]$ is given by their maximum: 
\begin{equation}
T[D] = \max_{P \in {\cal P}}  \ S_P. 
\end{equation} 
We remark that the variables $S_P$ 
are not independent, because two 
distinct paths in ${\cal P}$ may have nodes in common. 
Hence the PDF of $T[D]$ is given by 
\begin{equation} \label{FTD}
F_{T[D]}(t) = \prob(T[D] \leq t) = \prob(S_P \leq t \mbox{ \ for all  } P \in {\cal P}),
\end{equation} 
but no further simplification is in general possible.
To determine the pdf of $T[D]$, simply apply Equation~(\ref{rel}):
\begin{equation} 
f_{T[D]}(t) = (F_{T[D]})^{\prime}(t). 
\end{equation}

\bigskip

{\em Case 2.2: $D$ is a decision NDSAN}

\bigskip

In Figure~3(b), assume that the decision node is $d_i$. 
Then $\alpha_i = \eta-2$ and each node $source(D_k)$ 
is selected by $d_i$ with probability $p^i_k$, $k=2,3,\ldots,\eta-1$. 
Let $X_i$ be a random variable associated with $d_i$ 
in such a way that   
\begin{equation} \label{defX1}
X_i = \left\{ \begin{array}{c} 
T[D_2] \mbox{ with probability } p^i_2 \,; \\
\\
T[D_3] \mbox{ with probability } p^i_3 \,; \\
\vdots \\
T[D_{\eta-1}] \mbox{ with probability } p^i_{\eta-1} \,.
\end{array} \right. 
\end{equation}
Then, clearly, 
\begin{equation}
T[D] = T[D_1] + X_i + T[D_{\eta}].
\end{equation}

In order to proceed, note that the events $X_i = T[D_k]$, $2 \leq k \leq \eta-1$, 
are mutually disjoint, since they correspond to disjoint subdigraphs of $D$. 
We then have  
\begin{equation} \label{fX1}
f_{X_i}(t) = p^i_2 \ f_{T[D_2]}(t) + p^i_3 \ f_{T[D_3]}(t) 
+ \cdots + p^i_{\eta-1} \ f_{T[D_{\eta-1}]}(t)
\end{equation} 
and 
\begin{equation}
F_{X_i}(t) = p^i_2 \ F_{T[D_2]}(t) + p^i_3 \ F_{T[D_3]}(t) 
+ \cdots + p^i_{\eta-1} \ F_{T[D_{\eta-1}]}(t).
\end{equation} 
Thus, 
\begin{equation} \label{fTD}
f_{T[D]}(t) = (f_{T[D_1]} * f_{X_i} * f_{T[D_{\eta}]})(t) 
\end{equation}
and, by Equation~(\ref{rel}), 
\begin{equation} \label{FTD2}
F_{T[D]}(t) = \int_0^t f_{T[D]}(x) \ dx.  
\end{equation}












\bigskip
 
{\em Case 2.3: $D$ is a loop NDSAN}  

\bigskip

In Figure~4(b), assume that the loop node is $\ell_i$. 
For simplicity, assume also that $\beta_i=\beta$.
Recall that, 
at the $k$th passage through ${\ell}_i$, 
the execution flow returns to $source(D_2)$ 
with probability 
$q^i_k, k=1,\ldots,\beta$, where 
$q^i_{_{\beta}}=0$ and 
$\beta$ is the maximum number 
of consecutive passages allowed through $\ell_i$. 

Let $Z_k$ be the random variable standing for 
the total execution time of $k$ 
serial independent executions of $D_2$. Clearly, $Z_k$ 
is the sum of $k$ independent random variables, 
each one having distribution identical to that of $T[D_2]$. 
Therefore, $f_{Z_k}$ and $F_{Z_k}$ can once again be 
determined respectively by convolution and subsequent integration.     

Consider now a random variable $Y_i$ associated with 
$d_i$ and such that 
\begin{equation} \label{defY1}
Y_i = \left\{ \begin{array}{rl} 
0 &\mbox{with probability } 1-q^i_1 \,;\\
\\
Z_1 &\mbox{with probability } q^i_1(1-q^i_2) \,; \\
\\
Z_2 &\mbox{with probability } q^i_1 q^i_2 (1-q^i_3) \,; \\
 & \ \ \ \ \ \ \vdots \\
Z_{\beta-1} &\mbox{with probability } q^i_1 q^i_2 \cdots q^i_{_{\beta-1}} \, ,  
\end{array} \right. 
\end{equation} 
where the events $Y_i =0, \ Y_i=Z_1, \ \ldots, \ Y_i=Z_{\beta-1}$ are 
all mutually disjoint.
Then 
\begin{equation}
T[D] = T[D_1] + Y_i + T[D_3], 
\end{equation}
and the functions $f_{T[D]}$ and $F_{T[D]}$ can be 
obtained as in Case 2.2, since the definition of $Y_i$ in Equation~(\ref{defY1}) 
has the same structure as that of $X_i$ in Equation~(\ref{defX1}).   

\section{Obtaining an approximate distribution of the execution time}

Given an NDSAN $D$, obtaining 
the distribution and density  
functions of the target 
random variable $T[D]$ numerically may be an extremely costly computational task, 
even in simple cases. We refer the reader once again to the work 
by Leemis {\em et al.}~\cite{LDDMC06}, where even small networks 
are seen to need an elaborate mathematical analysis.


Our efforts are then directed toward seeking an 
approximate distribution of $T[D]$  
within some required confidence level. 
We base our approach on  
collecting a random sample formed by 
a suitable number $N$ of independent 
observations of $T[D]$. 
Let us denote such an approximate distribution by $F^{^N}_{T[D]}$. 
Once $F^{^N}_{T[D]}$ is obtained, a frequency histogram and 
an approximate density $f^{^N}_{T[D]}$ can be easily determined, 
as we discuss later.   

First, we present a simulation algorithm that, 
on input $D$, outputs a single observation  
$t$ of the sample space of $T[D]$. Next, we deal with the question of 
how many times the simulation algorithm must be repeated 
in order to obtain $F^{^N}_{T[D]}$ as required.  

\subsection{Simulation algorithm} 

The simulation algorithm is based on recursive references 
to subdigraphs, whose results are combined to obtain a single  
observation $t$ of $T[D]$.  
The basis of the recursion occurs when $D$ is a trivial NDSAN. 

For acyclic NDSANs (refer to item 2.1 in Definition 1 and to Figure~2(b)), a single 
observation of $T[D]$ is obtained as follows: (i)  
Observations $t_1, t_2, \ldots, t_{\eta}$ of 
$T[D_1], T[D_2], \ldots, T[D_{\eta}]$ are obtained recursively; (ii)
Denote by $C_D(t_1, t_2, \ldots, t_{\eta})$ 
the completion time of $D$ when $T[D_i]=t_i$, 
$1 \leq i \leq \eta$; the determination of $C_D(t_1, t_2, \ldots, t_{\eta})$ 
can be done by assigning weight $t_i$ to vertex $v_i$, 
$1 \leq i \leq \eta$, and then calculating the critical path of the 
resulting weighted digraph. 

The description of the simulation algorithm is as follows.

\newpage
{\sf  

\begin{tabbing}

Sample($D$)\\

1 \ \ if $D$ is a trivial NDSAN then\\

2  \hspace*{2cm} \= let $a_i$ be the single activity node of $D$\\

3 \> return a single observation of $T_i$\\

4 \ \ else if $D$ is an acyclic NDSAN then\\

5 \> let $D_1, D_2, \ldots D_{\eta}$ be NDSANs as in Figure~2(b)\\

6 \> return \ $C_D$( Sample($D_1$), Sample($D_2$)$,\ldots,$ Sample($D_{\eta}$) )\\

7 \ \ else if $D$ is a decision NDSAN then\\

8 \> let $D_1, D_2, \ldots D_{\eta}$ be NDSANs as in Figure~3(b)\\

9 \> let $d_i$ be the decision node of $D$\\
 
10 \> select $k$ from $\{2,3,\ldots,\eta-1\}$\\

11 \> return \ Sample($D_1$) \ $+$ \ Sample($D_k$) \ $+$ \ Sample($D_{\eta}$)\\

12 \ else if $D$ is a loop NDSAN then\\

13 \> let $D_1, D_2, D_3$ be NDSANs as in Figure~4(b)\\

14 \> let $\ell_i$ be the loop node of $D$\\

15 \> select $k$ from $\{0,1,\ldots,\beta_i-1\}$\\

16 \> $t_{loop} := 0$\\

17 \> repeat \= $k$ times\\ 

18 \> \> $t_{loop} := t_{loop} \ +$ Sample($D_2$)\\

19 \> return \ Sample($D_1$) \ $+$ \ $t_{loop}$ \ $+$  \ Sample($D_3$)\\

\end{tabbing}  

}

We assume that obtaining the single observation in Line $3$ 
can be done in constant time.   
We also assume that the selections in Lines $10$ and $15$ 
take constant time. Note that they   
are related to observations of the random variables $X_i$ 
and $Y_i$, respectively (see Equations~(\ref{defX1}) and~(\ref{defY1})).
Then they must be made according to 
the probabilities expressed there.  
Calculating $C_D$ in Line 6 takes 
$O(m)$ time. (The critical path can be determined by a depth-first 
search starting at $source(D)$.)

Overall, the time complexity of the algorithm is determined 
by the maximum number of nested loop NDSANs in $D$. Suppose 
that $D_1, D_2, \ldots, D_{\gamma}$ is the longest sequence of subdigraphs of 
$D$ such that: 

\begin{itemize}

\item[--] $D_k$ is a loop NDSAN, $1 \leq k \leq \gamma$;  

\item[--] $D_{k+1}$ is a proper subdigraph of $D_k$, $1 \leq k \leq \gamma-1$. 

\end{itemize}

Let $\bar{\beta}=\max\{\beta_i \mid 1 \leq i \leq n_{\ell}\}$. 
Then in each $D_k$ at most 
$\bar{\beta}-1$ consecutive iterations are performed. Hence, the 
worst-case time complexity of the algorithm is 
$O(\bar{\beta}^{\,\gamma} m)$. Although $\gamma=O(n)$ and $\bar{\beta}$ can 
be arbitrarily large, for most typical NDSANs 
the values of $\gamma$ and $\bar{\beta}$ are bounded by small constants. 
Thus the algorithm has, in practice, an $O(m)$ time complexity.   
  
\subsection{Repeated executions of the simulation algorithm} 

Since $F_{T[D]}$ is a continuous variable, we may resort to the same statistic
on $F^{^N}_{T[D]}$ as the Kolmogorov-Smirnov (KS) test.
We refer the reader to~\cite{Hoel} (Section~13.5) 
and to~\cite{Knuth} (Section~3.3.1) for more details on what follows.


Let $t_1, t_2, \ldots, t_{_N}$ be a random sample of $T[D]$,  
obtained by $N$ independent executions of the simulation algorithm. 
Define $F^{^N}_{T[D]}$ as 
\begin{equation} \label{FN}
F^{^N}_{T[D]}(x) = \frac{\mid \{t_i \mid t_i \leq x \} \mid }{N}.
\end{equation}   
The KS test is based on the difference between $F_{T[D]}(x)$ and 
$F^{^N}_{T[D]}(x)$. To measure this difference, we form the statistic 
\begin{equation} \label{KN}
K_{_N} = \ \underset{x \geq 0}{\sup} \ \mid  F^{^N}_{T[D]}(x) - F_{T[D]}(x) \mid
\end{equation}
(hereafter referred to as the KS statistic),
which may be visualized as the maximum distance (error), along the ordinate axis, 
between the plots of $F_{T[D]}(x)$ 
and $F^{^N}_{T[D]}(x)$ over the range of all possible $x$ values. 
It can be shown (see~\cite{Hoel}, p. 346) that the distribution 
of $K_{_N}$ does not depend on $F_{T[D]}$. 
As a consequence, $K_{_N}$ can be used as a nonparametric random variable for 
constructing a confidence band for $F_{T[D]}$. 

Let $K^{\varepsilon}_{_N}$ denote a value satisfying the relation 
\begin{equation} \label{PKN}
\prob(K_{_N} \leq K^{\varepsilon}_{_N}) =  1 - \varepsilon 
\end{equation}
for some $0 < \varepsilon < 1$.
Following Equations~(\ref{KN}) and (\ref{PKN}), we have: 
\begin{eqnarray}
1-\varepsilon &=&  \prob( \ \underset{x \geq 0}{\sup} \ 
\mid  F^{^N}_{T[D]}(x) - F_{T[D]}(x) \mid \ \leq \ K^{\varepsilon}_{_N})
\nonumber\\
&=&  \prob( \ \mid  F^{^N}_{T[D]}(x) - F_{T[D]}(x) \mid \ 
\leq \ K^{\varepsilon}_{_N} \ \mbox{for all} \ x \geq 0)
\nonumber\\
&=&  \prob( F^{^N}_{T[D]}(x) - K^{\varepsilon}_{_N} \leq F_{T[D]}(x)   
\leq F^{^N}_{T[D]}(x) + K^{\varepsilon}_{_N} \ \mbox{for all} \ x \geq 0).
\label{FNKN}
\end{eqnarray}
The last equality in Equation~(\ref{FNKN}) shows that the functions 
$F^{^N}_{T[D]}(x) - K^{\varepsilon}_{_N}$ and $F^{^N}_{T[D]}(x) + K^{\varepsilon}_{_N}$ 
yield a confidence band, with confidence level $1-\varepsilon$, for 
the unknown distribution function $F_{T[D]}(x)$.

Some of the values $K^{\varepsilon}_{_N}$ 
of the distribution of $K_{_N}$ are given in Table~1 
(see~\cite{Hoel}, p. 411). 
From Table~1 we have, for example, $K^{0.20}_{50}=0.15$. Thus 
\begin{equation}
\prob(K_{50} \leq K^{0.20}_{50}) = \prob(K_{50} \leq 0.15) = 1 - 0.20 = 0.80.
\end{equation}
That is, by repeating the simulation algorithm $N=50$ times, the probability 
that the error $K_{_N}$ is at most $0.15$ is $0.80$. 
More accurate results can be obtained by using the last row of Table~1. 
For example, by requiring a maximum error $0.02$ with confidence $95\%$, 
we have $\varepsilon=0.05$ and 
\begin{equation}
\prob(K_{_N} \leq K^{0.05}_{_N}) = \prob(K_{_N} \leq 0.02) = 0.95. 
\end{equation}
For large $N$, Table~1 gives us $K^{0.05}_{_N}=1.36/\sqrt{N}$. From 
$1.36/\sqrt{N}=0.02$ we conclude that $N=4624$ repeated executions 
of the simulation algorithm are needed in this case. 

\begin{table}[t]
\begin{center}
\caption{Some critical values $K^{\varepsilon}_{_N}$ for $K_N$.}
\begin{tabular}{|c|c|c|c|c|}
\hline
$ \ N \ $  & $\varepsilon=0.20$ & $\varepsilon=0.10$ 
& $\varepsilon=0.05$ & $\varepsilon=0.01$ \\ \hline
10 & 0.32 & 0.37 & 0.41 & 0.49\\
20 & 0.23 & 0.26 & 0.29 & 0.36\\
30 & 0.19 & 0.22 & 0.24 & 0.29\\
40 & 0.17 & 0.19 & 0.21 & 0.25\\
50 & 0.15 & 0.17 & 0.19 & 0.23\\
large & \ \ \ $1.07/\sqrt{N}$ \ \ \ & \ \ \ $1.22/\sqrt{N}$ \ \ \ 
& \ \ \ $1.36/\sqrt{N}$ \ \ \ & \ \ \ $1.63/\sqrt{N}$ \ \ \ \\
\hline  
\end{tabular}
\end{center}
\end{table}

We can summarize the application of the KS statistic as follows.

\begin{enumerate} 

\item Stipulate the maximum error $e$ and the confidence level $c$. 

\item Set $\varepsilon=1-c$ and determine from Table~1 
the value of $N$ for which  
$K^{\varepsilon}_{_N} \approx e$. 

\item Run the simulation algorithm $N$ times and obtain a random 
sample $t_1, t_2, \ldots, t_N$.

\item Let $F^{^N}_{T[D]}$ be as in Equation~(\ref{FN}). 

\item If needed, an approximate density $f^{^N}_{T[D]}$ 
can be determined as follows, assuming $t_1 \leq t_2 \leq \cdots \leq t_N$.
For some step value $\delta > 0$, let   
\begin{equation} \label{appdens}
f^{^N}_{T[D]}(t_{_{1+k\delta}})=
\frac{F^{^N}_{T[D]}(t_{_{1+k\delta}})-F^{^N}_{T[D]}(t_{_{1+(k-1)\delta}})}{t_{_{1+k\delta}}- t_{_{1+(k-1)\delta}}}, \ \ \ k=1,2,\ldots, \lfloor N/\delta \rfloor-1.
\end{equation} 
For instance, for $\delta=25$ we compute the values 
\[f^{^N}_{T[D]}(t_{_{26}})=\frac{F^{^N}_{T[D]}(t_{_{26}})-F^{^N}_{T[D]}(t_{_1})}{t_{_{26}}- t_{_1}} \ , \  
  f^{^N}_{T[D]}(t_{_{51}})=\frac{F^{^N}_{T[D]}(t_{_{51}})-F^{^N}_{T[D]}(t_{_{26}})}{t_{_{51}}- t_{_{26}}} \ , \] 
and so on. (We remark that better, nonparametric methods are available, as
explained in \cite{silverman}, for example.)

\end{enumerate}

\section{Computational experiments}


\subsection{A typical development process}

Figure~5 shows a simple, yet typical, development 
process represented by a NDSAN $D$ with $S_a=\{a_1,\ldots,a_{27}\}$, 
$S_b=\{b_1,\ldots,b_8\}$, $S_d=\{d_1\}$, and $S_{\ell}=\{\ell_1, \ldots, \ell_7\}$. 

\begin{figure}[p]
\label{exemplo1}
\epsfxsize=16cm
\centerline{\epsfbox{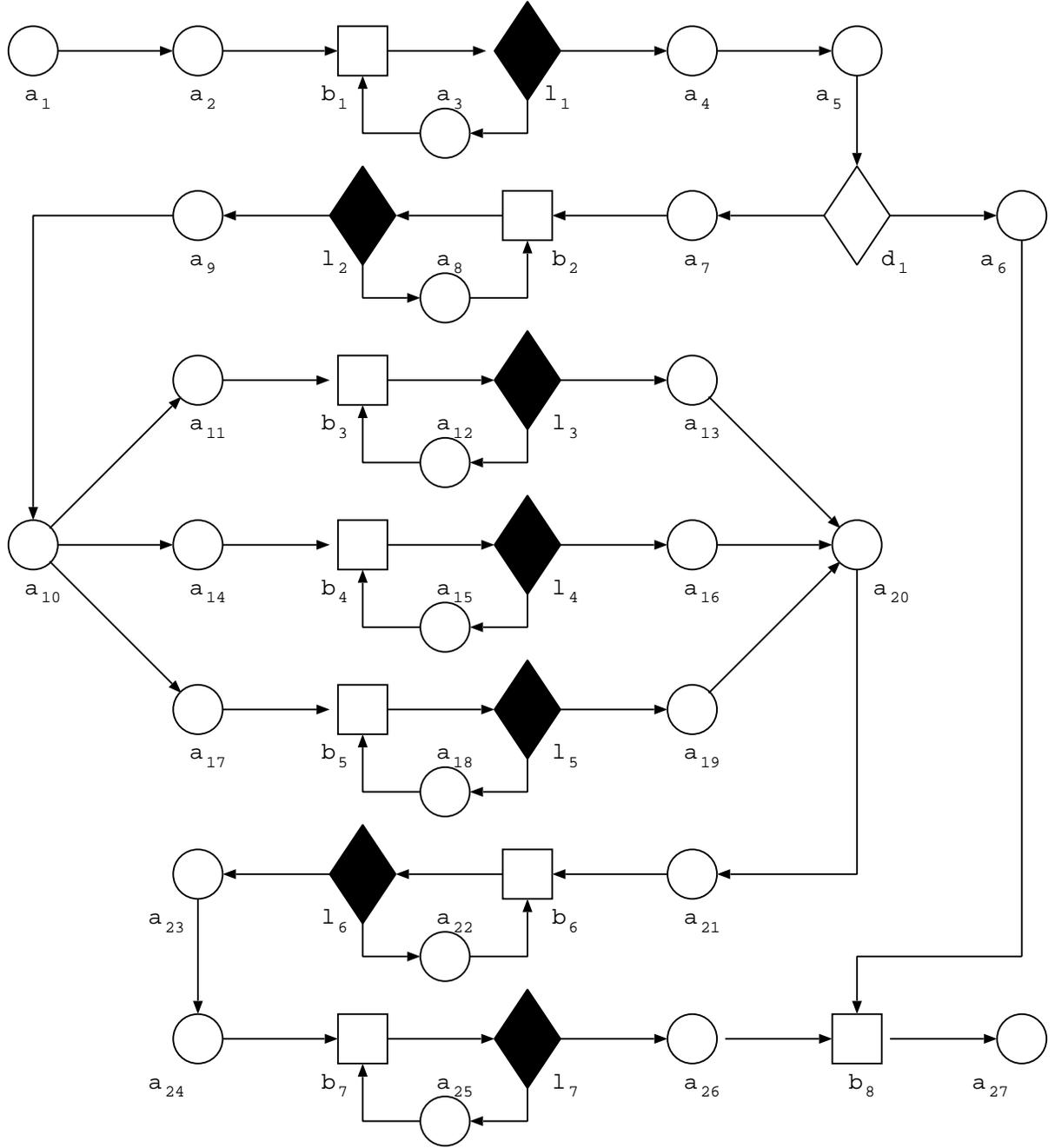}}
\caption{An NDSAN representing a development process.}
\end{figure}

Table~2 describes the activity nodes, whose durations are expressed in days.
Here, all $T_i$'s follow {\em triangular densities}, 
which are suitable for describing single activities of a business 
or industrial process~\cite{triang}. The pdf $f_X$ of a triangular variable  
$X$ with parameters $x_1 < x_2 < x_3$ is given by:
\begin{equation}
f_X(x)=\left\{ \begin{array}{ll} 
0, & x < x_1;\\
\\
\frac{y_{_0}}{x_2-x_1}(x-x_1), & x_1 \leq x < x_2;\\
\\
\frac{y_{_0}}{x_3-x_2}(x_3-x), & x_2 \leq x < x_3;\\
\\
0, & x \geq x_3, 
\end{array} \right.
\end{equation}
where $y_0=\frac{2}{x_3-x_1}$.    
Table~3 shows the probabilities associated with the decision node $d_1$,  
Table~4 those associated with the loop nodes $\ell_1$ through $\ell_7$.

\begin{table}[p]
\begin{center}
\caption{Activity nodes of the NDSAN of Figure~5.}
\begin{tabular}{|c|l|c|}
\hline
Node & Description & Density parameters\\ 
\hline\hline
$a_1$ & requirement analysis & $2,\ 4, \ 5$\\
$a_2$ & contract negotiation & $1, \ 2.5, \ 3.5$\\
$a_3$ & renegotiation & $1, \ 1.5, \ 2$\\
$a_4$ & contract conclusion & $0.5, \ 1, \ 1.5$ \\
$a_5$ & contract presentation & $0.5, \ 1, \ 1.5$\\
$a_6$ & project abandonment& $0.5, \ 1, \ 1.5$\\
$a_7$ & system analysis & $4, \ 8, \ 12$\\
$a_8$ & system analysis refinement & $0.5, \ 2, \ 3$ \\
$a_9$ & system analysis conclusion & $0.5, \ 1, \ 1.5$ \\
$a_{10}$ & division into modules & $0.5, \ 1, \ 1.5$\\
$a_{11}$ & 1st module implementation & $4, \ 6, \ 12$ \\
$a_{12}$ & 1st module refinement & $1, \ 2, \ 3$\\
$a_{13}$ & 1st module conclusion & $0.5, \ 1, \ 1.5$\\
$a_{14}$ & 2nd module implementation & $4, \ 6, \ 12$\\
$a_{15}$ & 2nd module refinement & $1, \ 2, \ 3$\\
$a_{16}$ & 2nd module conclusion & $0.5, \ 1, \ 1.5$\\
$a_{17}$ & 3rd module implementation & $4, \ 6, \ 12$ \\
$a_{18}$ & 3rd module refinement & $1, \ 2, \ 3$\\
$a_{19}$ & 3rd module conclusion & $0.5, \ 1, \ 1.5$\\
$a_{20}$ & module integration & $0.5, \ 1.5, \ 3$\\
$a_{21}$ & integration test & $1, \ 3.5, \ 4$\\
$a_{22}$ & error fixing & $0.5, \ 1, \ 1.5$\\
$a_{23}$ & product deployment & $0.5, \ 1, \ 1.5$\\
$a_{24}$ & client test & $2, \ 4, \ 6$\\
$a_{25}$ & error fixing & $0.5, \ 1, \ 1.5$\\
$a_{26}$ & production dispatch & $0.5, \ 1, \ 1.5$\\
$a_{27}$ & project documentation & $0.5, \ 1, \ 1.5$\\
\hline  
\end{tabular}
\end{center}
\end{table} 

\begin{table}[p]
\begin{center}
\caption{Probabilities associated with the decision node $d_1$ in Figure~5.}
\begin{tabular}{|c|l|c|c|c|}
\hline
Node & Description & Outcome & Next activity & Probability\\ 
\hline\hline
$d_1$ & contract accepted? & {\tt yes} & $a_7$ & 55\%\\
& & {\tt no} & $a_6$ & 45\%\\ 
\hline
\end{tabular}
\end{center}
\end{table} 

\begin{table}[p]
\begin{center}
\caption{Probabilities associated with the loop nodes in Figure~5.}
\begin{tabular}{|c|l|c|c|c|c|c|}
\hline
Node & Description & Outcome & Next activity & 1st iter. & 2nd iter. & 3rd iter.\\ 
\hline\hline
$\ell_1$ & negotiation finished? & {\tt yes} & $a_4$ & 50\% & 80\% & 100\%\\
& & {\tt no} & $a_3$ & 50\% & 20\% & 0\%\\ 
\hline
$\ell_2$ & use cases approved? & {\tt yes} & $a_9$ & 10\% & 50\% & 100\%\\
& & {\tt no} & $a_8$ & 90\% & 50\% & 0\%\\
\hline
$\ell_3$ & 1st module passed? & {\tt yes} & $a_{13}$ & 20\% & 50\% & 100\%\\
& & {\tt no} & $a_{12}$ & 80\% & 50\% & 0\%\\
\hline
$\ell_4$ & 2nd module passed? & {\tt yes} & $a_{16}$ & 20\% & 50\% & 100\%\\
& & {\tt no} & $a_{15}$ & 80\% & 50\% & 0\%\\
\hline
$\ell_5$ & 3rd module passed? & {\tt yes} & $a_{19}$ & 20\% & 50\% & 100\%\\
& & {\tt no} & $a_{18}$ & 80\% & 50\% & 0\%\\
\hline
$\ell_6$ & integration passed? & {\tt yes} & $a_{23}$ & 60\% & 80\% & 100\%\\
& & {\tt no} & $a_{22}$ & 40\% & 20\% & 0\%\\
\hline
$\ell_7$ & client test passed? & {\tt yes} & $a_{26}$ & 20\% & 50\% & 100\%\\
& & {\tt no} & $a_{25}$ & 80\% & 50\% & 0\%\\
\hline   
\end{tabular}
\end{center}
\end{table}

If we require a maximum error of $2\%$ with confidence $95\%$, 
the KS statistic yields $K^{0.05}_{_N}=1.36/\sqrt{N}$ (see Table~1). From  
$1.36/\sqrt{N}=0.02$, we conclude that $N=4624$ repeated executions of 
{\sf Sample}($D$) are required. 
Each of these executions can be represented by a tree 
of recursive calls, as follows. 
Let $D_i$ be the trivial NDSAN consisting of the activity node $a_i$, $1 \leq i \leq 27$, 
and, for $i < j$, let $D_{i,j}$ be the NDSAN defined 
as the maximal connected induced subdigraph 
$D'$ of $D$ satisfying $source(D')=a_i$ and $sink(D')=a_j$. 
Figure~6 depicts the tree of recursive calls. For example, 
$D_{5,27}$ is a decision NDSAN, and in order to obtain a single observation of $T[D_{5,27}]$ 
we first recursively obtain observations of 
$T[D_5], T[D_6], T[D_{7,26}]$, and $T[D_{27}]$. 

\begin{figure}[p]
\label{tree}
\epsfxsize=16cm
\centerline{\epsfbox{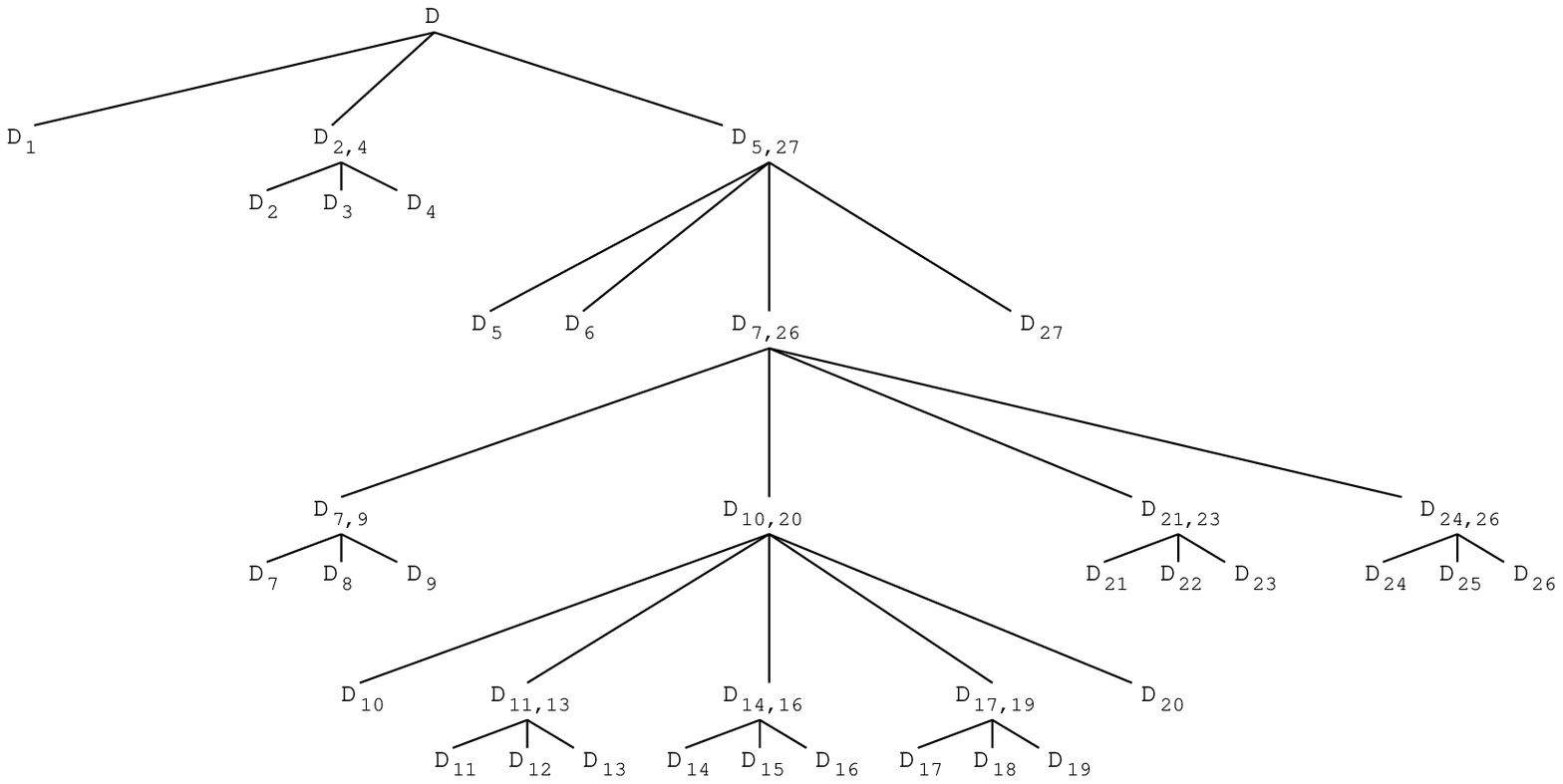}}
\caption{Recursive calls invoked by {\sf Sample}($D$); $D$ is the NDSAN of Figure~5.}
\end{figure}

The frequency histogram of the resulting sample of $T[D]$ is 
shown in Figure~7 for 1-wide bins. Each bin is an interval of the form 
$(a,b]$ and abscissae in the figure give the values of $b$.
The histogram suggests that $f_{T[D]}$ follows a bimodal pattern. 
Figure~7 also shows the fitting curve 
\begin{equation}
f_1(x) = 2115 \ \mbox{lognorm}(2.379610, 0.125138, x) + 
2509 \ \mbox{lognorm}(3.853650, 0.072067, x), 
\end{equation}
where $\mbox{lognorm}(\mu, \sigma, x)$ is the 
density function of the {\em log normal distribution} \cite{mathworld}
with parameters $\mu$ (the {\em scale} parameter) and $\sigma$ 
(the {\em shape} parameter):
\begin{equation}
\mbox{lognorm}(\mu, \sigma, x) =
\frac{1}{\sigma\sqrt{2\pi}x}
e^{-(\ln x-\mu)^2/2\sigma^2}.
\end{equation}
The function $f_1(x)$ is therefore proportional to
the sum of two densities, the former yielding positive values over the range $(7,16]$, 
the latter over $(37,60]$.  

\begin{figure}[t]
\epsfxsize=15cm
\centerline{\epsfbox{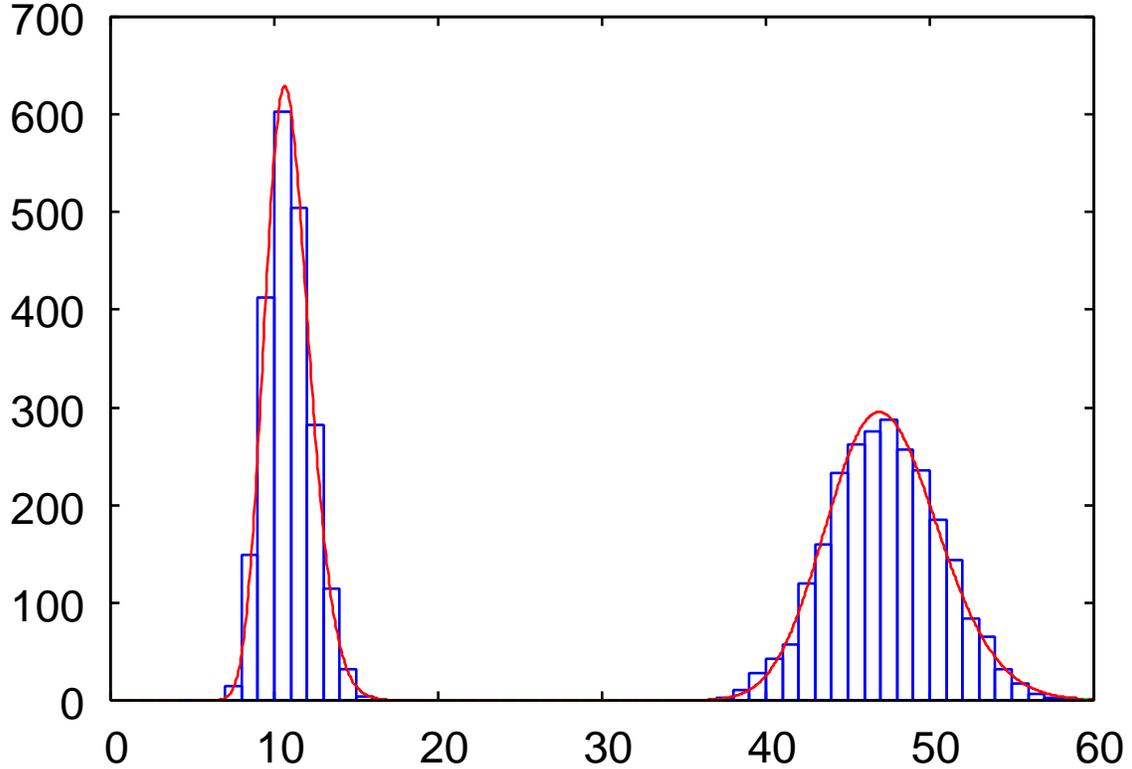}}
\caption{Fitting curve drawn on the frequency histogram of $T[D]$ for $N=4624$ and 1-wide bins; 
$D$ is the NDSAN of Figure~5.}
\end{figure}

The approximate $F^{^N}_{T[D]}$ 
and $f^{^N}_{T[D]}$  
are shown in Figures~8 and~9 respectively, the latter with 
$\delta=25$ in Equation~(\ref{appdens}).

\begin{figure}[p]
\epsfxsize=15cm
\centerline{\epsfbox{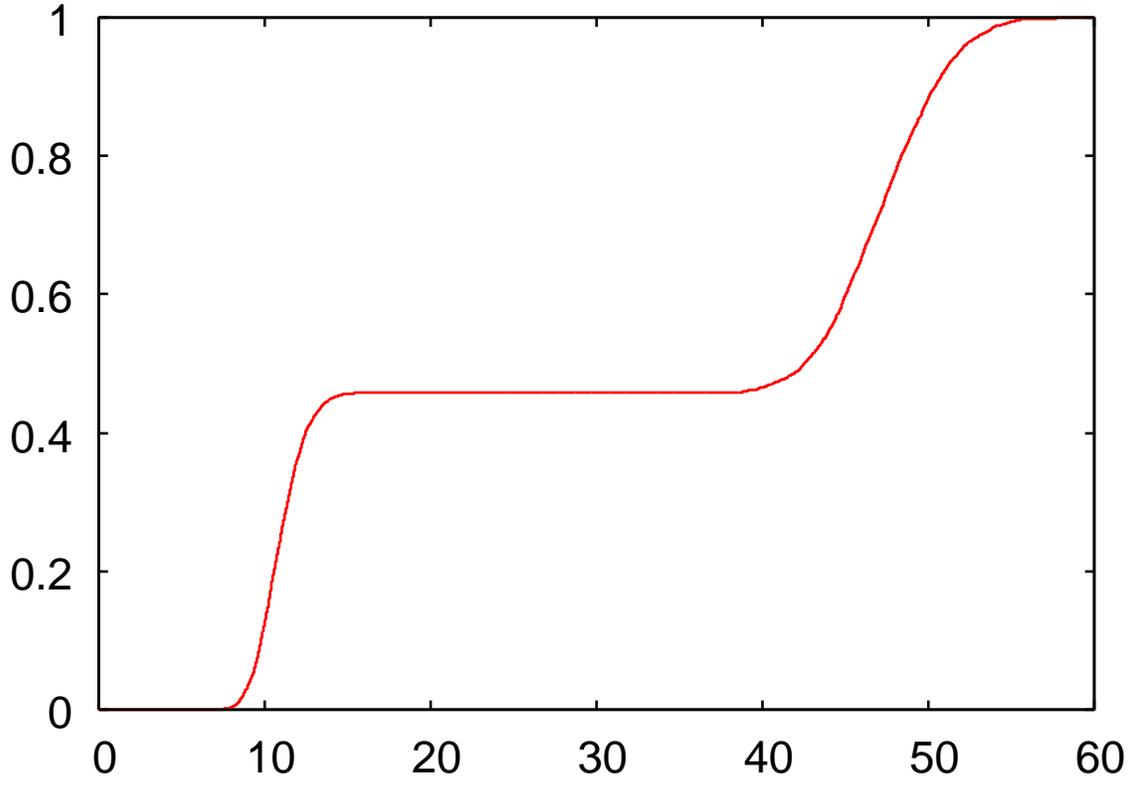}}
\caption{Approximate distribution $F^{^N}_{T[D]}$ for $N=4624$; 
$D$ is the NDSAN of Figure~5.}
\end{figure}

\begin{figure}[p]
\epsfxsize=15cm
\centerline{\epsfbox{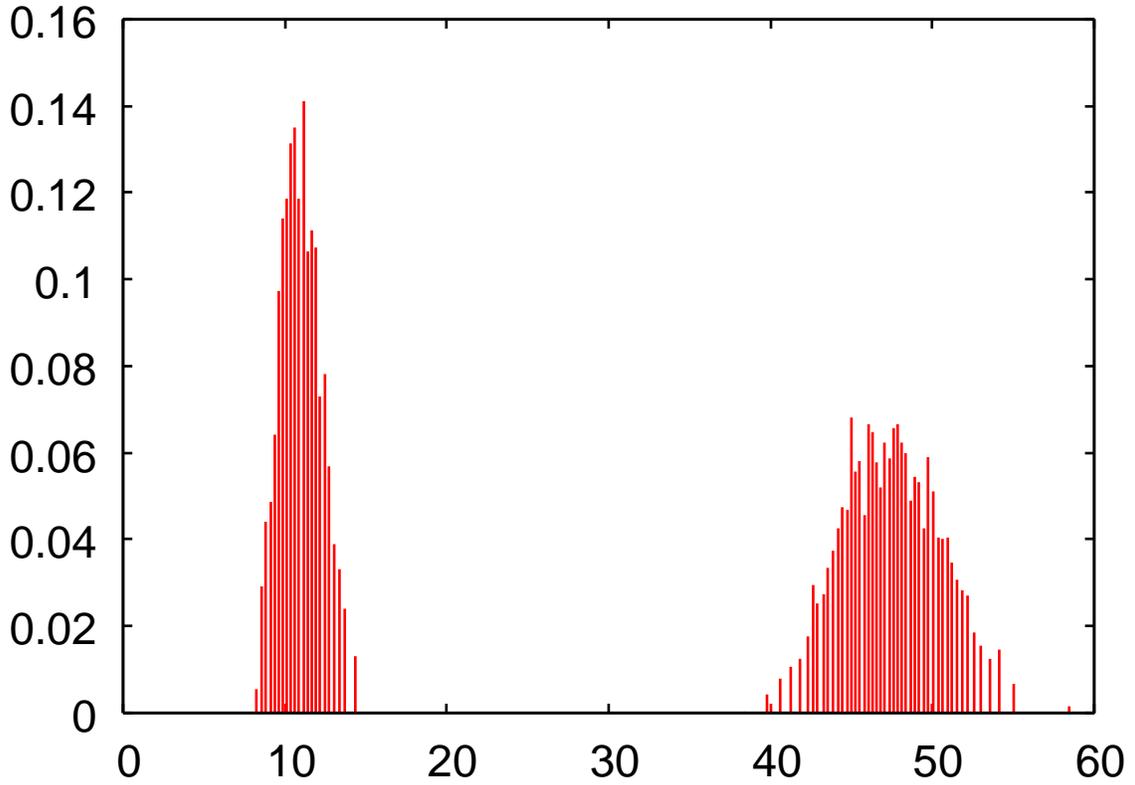}}
\caption{Approximate density $f^{^N}_{T[D]}$ for $N=4624$ and $\delta=25$; 
$D$ is the NDSAN of Figure~5.}
\end{figure}

\subsection{A paper reviewing process}

Figure~10 shows an NDSAN $D$ representing the typical peer-review process of scientific 
publishing. 
Table~5 describes the activity nodes, whose durations are once again   
expressed in days. The $T_i$'s follow {\em truncated normal distributions}. In the 
third column of Table~5, each line shows a pair $\mu_i, \sigma_i^2$,  
standing for the mean and the variance of $T_i$, respectively. 
Each $T_i$ is restricted to lie in the range $[\mu_i-3\sigma_i, \mu_i+3\sigma_i]$. 
Table~6 shows the probabilities associated with the decision node $d_1$, 
Table~7 the probabilities associated with the loop nodes $\ell_1$ and $\ell_2$.

\begin{figure}[p]
\label{exemplo2}
\epsfxsize=16cm
\centerline{\epsfbox{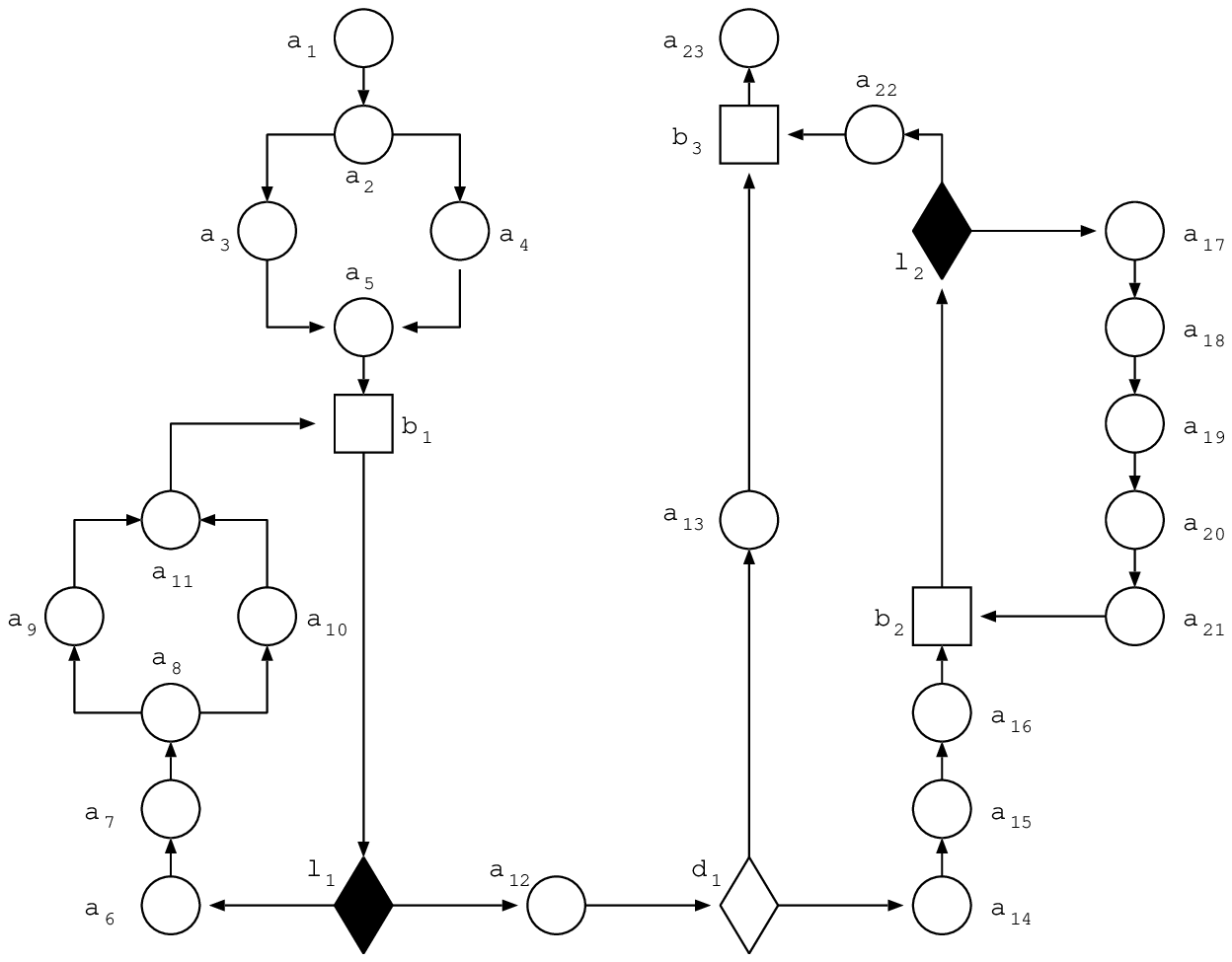}}
\caption{An NDSAN representing a paper reviewing process.}
\end{figure}

\begin{table}[p]
\begin{center}
\caption{Activity nodes of the NDSAN in Figure~10.}
\begin{tabular}{|c|l|c|}
\hline
Node & Description & Mean, variance\\ 
\hline\hline
$a_1$    & authors submit paper & 1, 0.1\\
$a_2$    & editor sends paper to referees 1 and 2 & 1, 0.1\\
$a_3$    & referee 1 processes the paper & 90, 45\\
$a_4$    & referee 2 processes the paper & 90, 45\\
$a_5$    & editor processes reports & 2, 0.2\\
$a_6$    & editor sends reports to authors & 1, 0.1\\
$a_7$    & authors perform modifications & 14, 7\\
$a_8$    & editor sends revised version to referees 1 and 2 & 1, 0.1\\
$a_9$    & referee 1 processes revised version & 14, 7\\
$a_{10}$ & referee 2 processes revised version & 14, 7\\
$a_{11}$ & editor processes new reports & 2, 0.2\\
$a_{12}$ & editor checks agreement of reports & 1, 0.1\\
$a_{13}$ & editor makes final decision based on two reports & 2, 0.2\\
$a_{14}$ & editor sends paper to referee 3 & 1, 0.1\\
$a_{15}$ & referee 3 processes the paper & 90, 45\\
$a_{16}$ & editor processes report of referee 3 & 2, 0.2\\
$a_{17}$ & editor sends report of referee 3 to authors & 1, 0.1\\
$a_{18}$ & authors perform modifications & 14, 7\\
$a_{19}$ & editor sends revised version to referee 3 & 1, 0.1\\
$a_{20}$ & referee 3 processes revised version  & 14, 7\\
$a_{21}$ & editor processes new report of referee 3 & 2, 0.2\\
$a_{22}$ & editor makes final decision based on three reports & 2, 0.2\\
$a_{23}$ & editor sends final result to authors & 1, 0.1\\

\hline  
\end{tabular}
\end{center}
\end{table} 

\begin{table}[p]
\begin{center}
\caption{Probabilities associated with the decision node $d_1$ in Figure~10.}
\begin{tabular}{|c|l|c|c|c|}
\hline
Node & Description & Outcome & Next activity & Probability\\ 
\hline\hline
$d_1$ & referees agree? & {\tt yes} & $a_{13}$ & 75\%\\
& & {\tt no} & $a_{14}$ & 25\%\\ 
\hline
\end{tabular}
\end{center}
\end{table} 

\begin{table}[p]
\begin{center}
\caption{Probabilities associated with the loop nodes in Figure~10.}
\begin{tabular}{|c|l|c|c|c|c|c|}
\hline
Node & Description & Outcome & Next activity & 1st iter. & 2nd iter. & 3rd iter.\\ 
\hline\hline
$\ell_1$ & no need of modifications? & {\tt yes} & $a_{12}$ & 81\% & 98\% & 100\%\\
& & {\tt no} & $a_{6}$ & 19\% & 2\% & 0\%\\ 
\hline
$\ell_2$ & no need of modifications? & {\tt yes} & $a_{22}$ & 90\% & 99\% & 100\%\\
& & {\tt no} & $a_{17}$ & 10\% & 1\% & 0\%\\
\hline   
\end{tabular}
\end{center}
\end{table} 

For the same $2\%$ error and $95\%$ confidence as above, we give the results
from  $N=4624$ repeated executions of {\sf Sample}($D$) in 
Figures~11 through~13. These figures show, respectively, the 
fitting curve $f_2(x)= 4624 \ \mbox{lognorm}(4.965323, 0.421285, x)$ 
drawn on the frequency histogram of $T[D]$ for 1-wide bins, 
the approximate distribution of $T[D]$, and  
the approximate density of $T[D]$ (with 
$\delta=25$ in Equation~(\ref{appdens})). 


\begin{figure}[t]
\epsfxsize=15cm
\centerline{\epsfbox{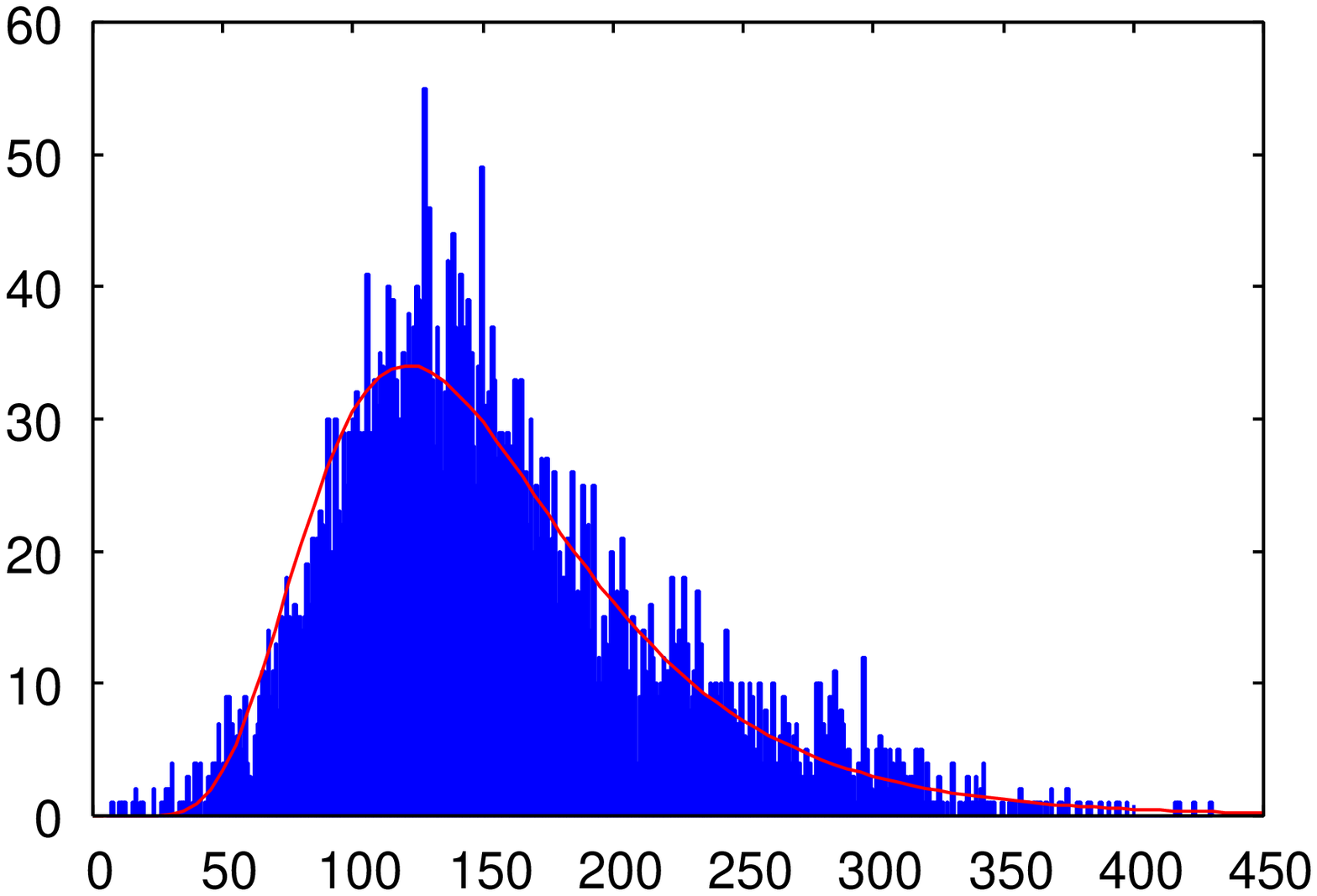}}
\caption{Fitting curve drawn on the 
frequency histogram of $T[D]$ for $N=4624$ and 1-wide bins;
$D$ is the NDSAN of Figure~10.}
\end{figure}

\begin{figure}[p]
\epsfxsize=15cm
\centerline{\epsfbox{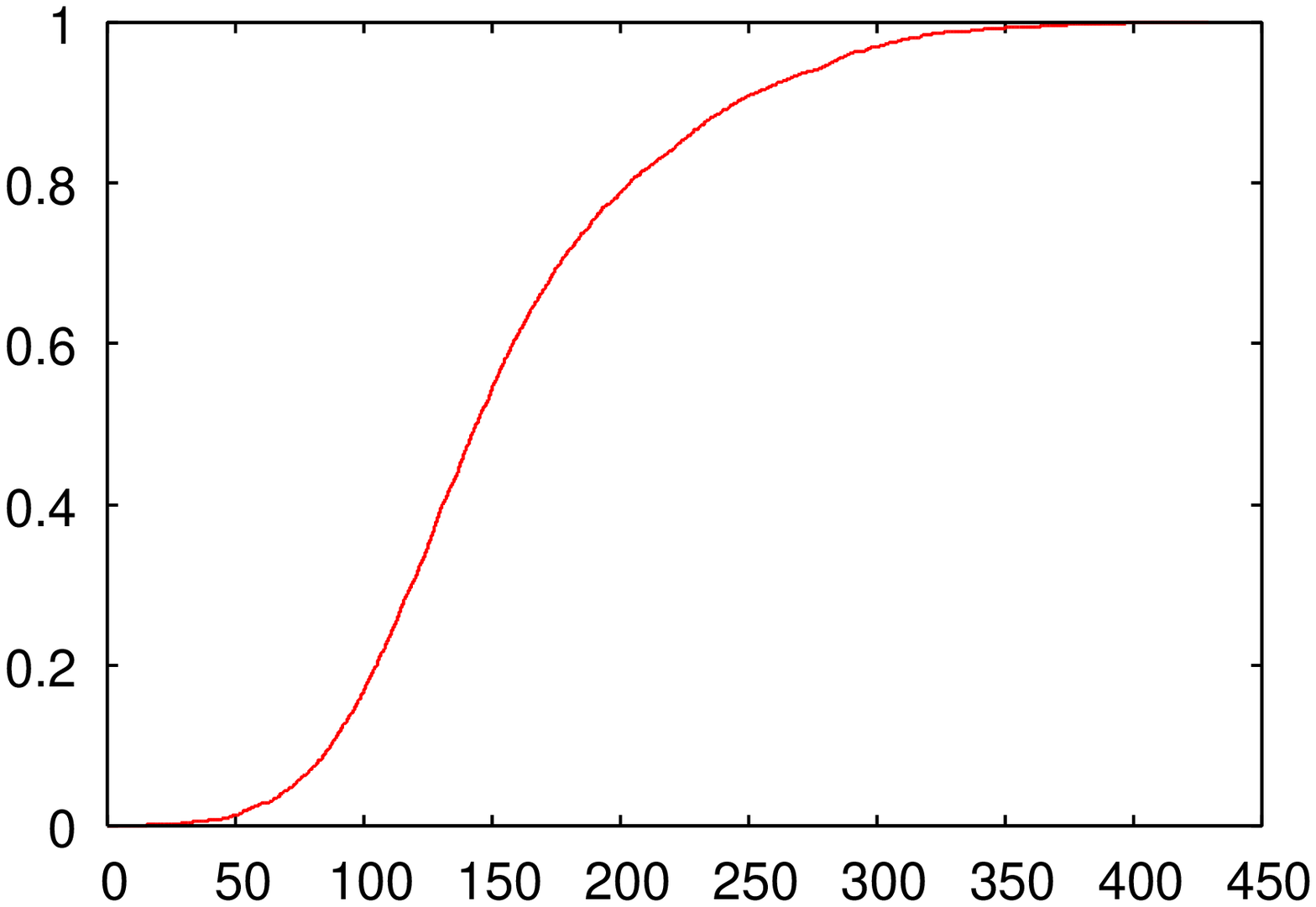}}
\caption{Approximate distribution $F^{^N}_{T[D]}$ for $N=4624$; 
$D$ is the NDSAN of Figure~10.}
\end{figure}

\begin{figure}[p]
\epsfxsize=15cm
\centerline{\epsfbox{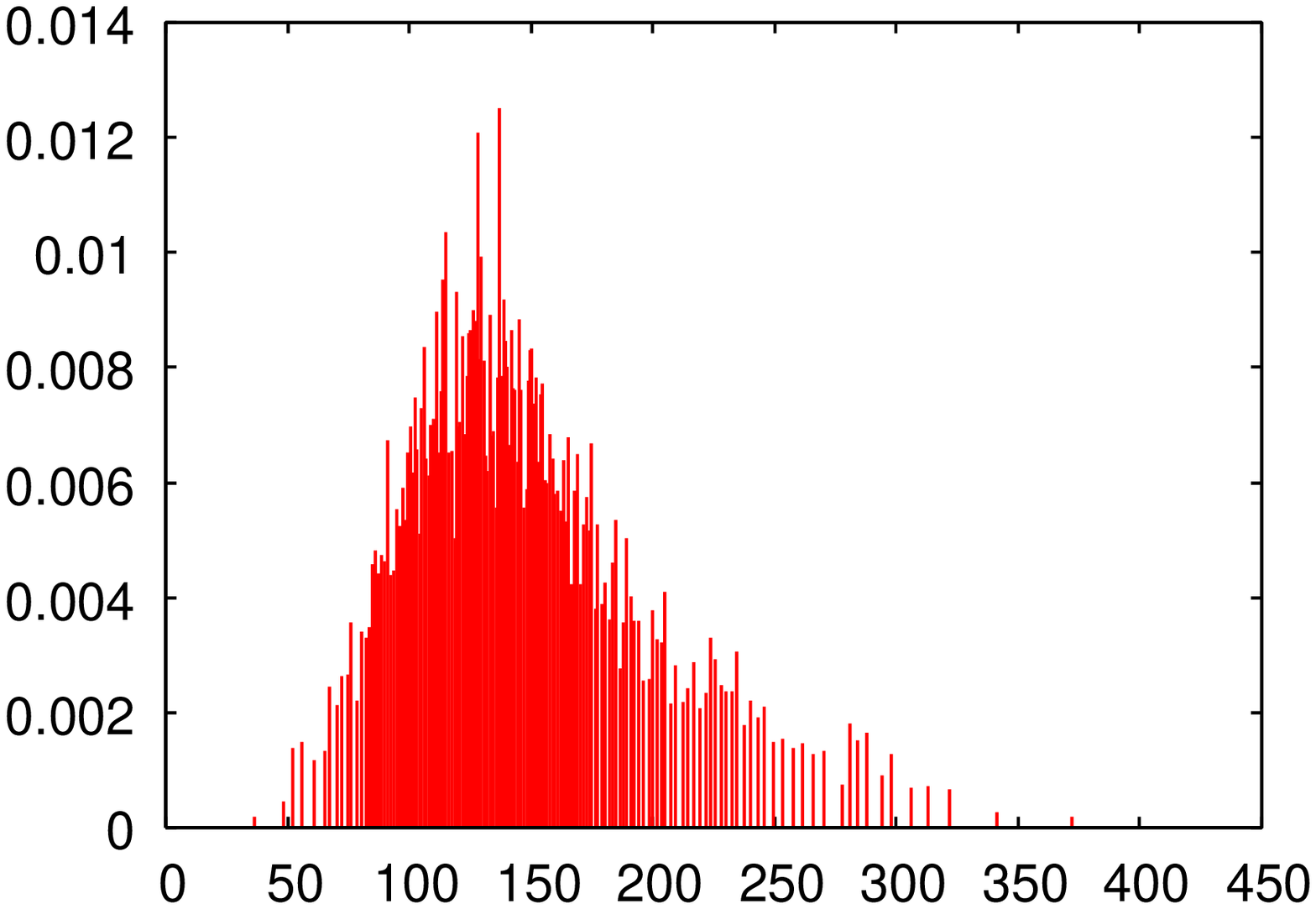}}
\caption{Approximate density $f^{^N}_{T[D]}$ for $N=4624$ and $\delta=25$; 
$D$ is the NDSAN of Figure~10.}
\end{figure}

\section{Ongoing work}

The introduction of the constraint that each activity node requires certain amounts of finitely available resources to execute gives raise to the so-called {\em activity networks with constrained resources}.  The problem associated with such networks is known as RCPSP (Resource-Constrained Project Scheduling Problem)~\cite{D99}. The RCPSP has many variations, but even the deterministic RCPSP with fixed activity durations is NP-hard~\cite{BLR83}. 

Resource-Constrained NDSANs (RCNDSANs) combine stochastic activity durations, non\-de\-ter\-minism, and constrained resources. We are currently targeting the simulation algorithm of RCNDSANs, based on iterating the combination of two phases as many times as necessary for accuracy. The first phase is responsible for obtaining a non-stochastic, deterministic instance of the input RCNDSAN, by selecting one of its possible execution paths. (Here, the term ``path'' stands for a plausible non-stochastic, deterministic scenario: a network represented by a directed acyclic graph with fixed topology and fixed activity durations.) The second phase consists of employing a heuristic procedure for the solution of the deterministic RCPSP. The repeated execution of ``path selection'' combined with ``scheduling heuristics'' will generate close approximations to the probability distribution of the variables under analysis.

We remark that our simulation algorithms turn out to be low-cost tools for the identification of the factors that most strongly influence completion time. After a simulation round, if needed, changes in the structure of the NDSAN/RCNDSAN under analysis can be proposed in order to improve its performance. Several simulation rounds may be rapidly performed until the desired efficiency is actually achieved.

\end{document}